\documentstyle[psfig]{lamuphys}
\makeatletter
\let\chapter\hid@chapter
\makeatother
\def\aeta{A\&A }
\def\aetal{A\&AL }
\def\aetas{A\&AS }
\def\apj{ApJ }
\def\apjs{ApJS }
\def\aj{AJ }
\def\mn{MNRAS }

\def\apjl{ApJL \rm}
\begin{document}
\pagenumbering{arabic}
\title{Clustering of absorption line systems}

\author{Patrick\,Petitjean\inst{1,2}}

\institute{Institut d'Astrophysique de Paris -- CNRS\\
98bis, Boulevard Arago\\
F-75014 Paris, France
\and
DAEC, URA CNRS 173, Observatoire de Paris--Meudon\\
F-92195 Meudon, France}

\maketitle

\begin{abstract}
Absorption line systems are luminosity unbiased tracers of the 
spatial distribution of baryons over most of the history of the Universe.
I review the importance of studying the clustering properties 
of the absorbers and the impact of VLT in this subject.
The primary aim of the project is to track the evolution of the structures
of the Universe back in time.
\end{abstract}
\section{Introduction}
Evolution of large scale structures of the Universe is one of the most
important issues of modern cosmology.
QSO absorption line systems probe
material lying on the line of sight to quasars over a large
redshift range (0~$<$~$z$~$<$~5). The systems where metal lines
are detected have been recognized to be
associated with haloes of galaxies (e.g. Bergeron \& Boiss\'e 1991, Steidel
1993). Those with very low metal content (the Ly$\alpha$ forest),
are generally believed to probe intergalactic gas. Part of this gas could
be associated with galaxies however, 
but how is yet unclear (e.g. Lanzetta et al. 1994, Le~Brun et al. 
1995, Charlton et al. 1995). In any case, the absorption line systems can 
be used as luminosity unbiased tracers of the spatial distribution
of baryons over most of the history of the Universe.
To do so background sources at small projected distances on the sky should 
be observed 
to study correlation between absorptions detected along different lines of 
sight. When the sources are very close (for example two images of the same
gravitationally lensed quasar), the lines of sight probe the same clouds
and information on the perpendicular sizes and the internal structure
of the clouds can be derived. With sources further away
from each others, one can study the correlation length of the clouds.\par
This has been recognized for over a decade (e.g. Shaver \& Robertson 1983,
Robertson \& Shaver 1983).
Observations of QSO pairs with projected separations from a few arcseconds to
a few arcminutes yield interesting constraints on the size,
physical structure and kinematics of
galatic haloes, clusters and filaments. Indeed, new
constraints have been obtained very recently on the extent
of the Ly$\alpha$ complexes perpendicular to the line of sight
at high (Smette et al. 1992, 1995, Bechtold et al. 1994, Dinshaw et al.
1994) and intermediate
(Dinshaw et al. 1995) redshifts indicating that they could
have sizes larger than 300~kpc. Such sizes are more indicative of
a correlation length than of real cloud sizes (Rauch \& Haenelt 1996).
This is consistent with the picture that the Ly$\alpha$ gas traces
the potential wells of dark matter filamentary structures (Cen et al.
1994, Petitjean et al. 1995, M\"ucket et al. 1996, Hernquist et al. 1996,
Miralda-Escud\'e et al. 1996).
Large scale clustering of C~{\sc iv} systems (Heisler et al. 1989;
Foltz et al. 1993) or damped systems (Francis \& Hewett 1993, Wolfe 1993)
have also been detected recently.
The advent of 10m-class telescopes
will boost this field since observation of faint QSOs in the same field
will allow 3-D mapping of the baryonic content of the Universe via 
absorption line systems (Petitjean 1995).\par
\begin{figure}
\centerline{\vbox{
\psfig{figure=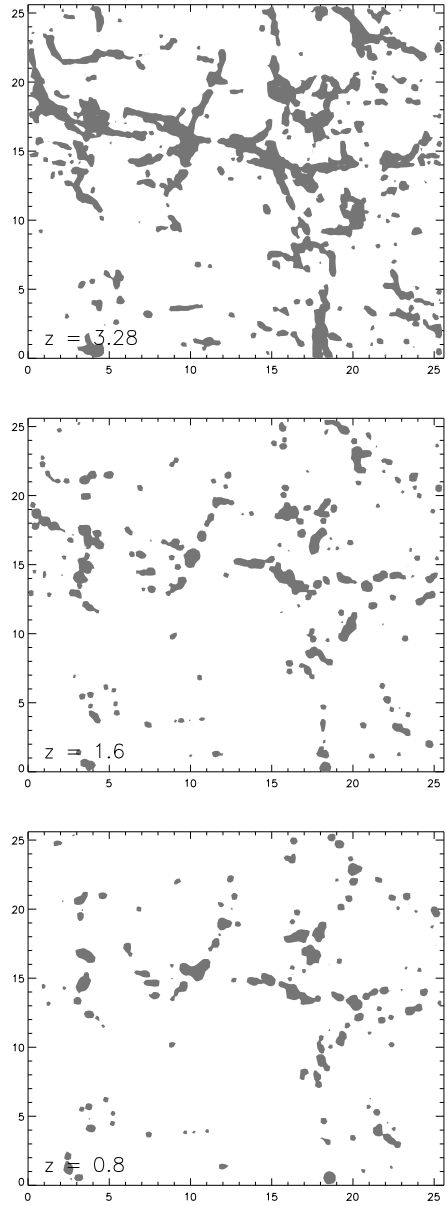,height=16.5cm}
}}
\caption{Spatial distribution of Ly$\alpha$ clouds in a 2 Mpc slice of
a (25 Mpc)$^3$ simulation box at redshifts $z$ = (3.28, 1.6, 0.8) 
respectively. Gray contours show regions where 
$N$(H~{\sc i})~$>$~10$^{13}$~cm$^{-2}$ through the box.}
\end{figure}
\begin{figure}
\centerline{\vbox{
\psfig{figure=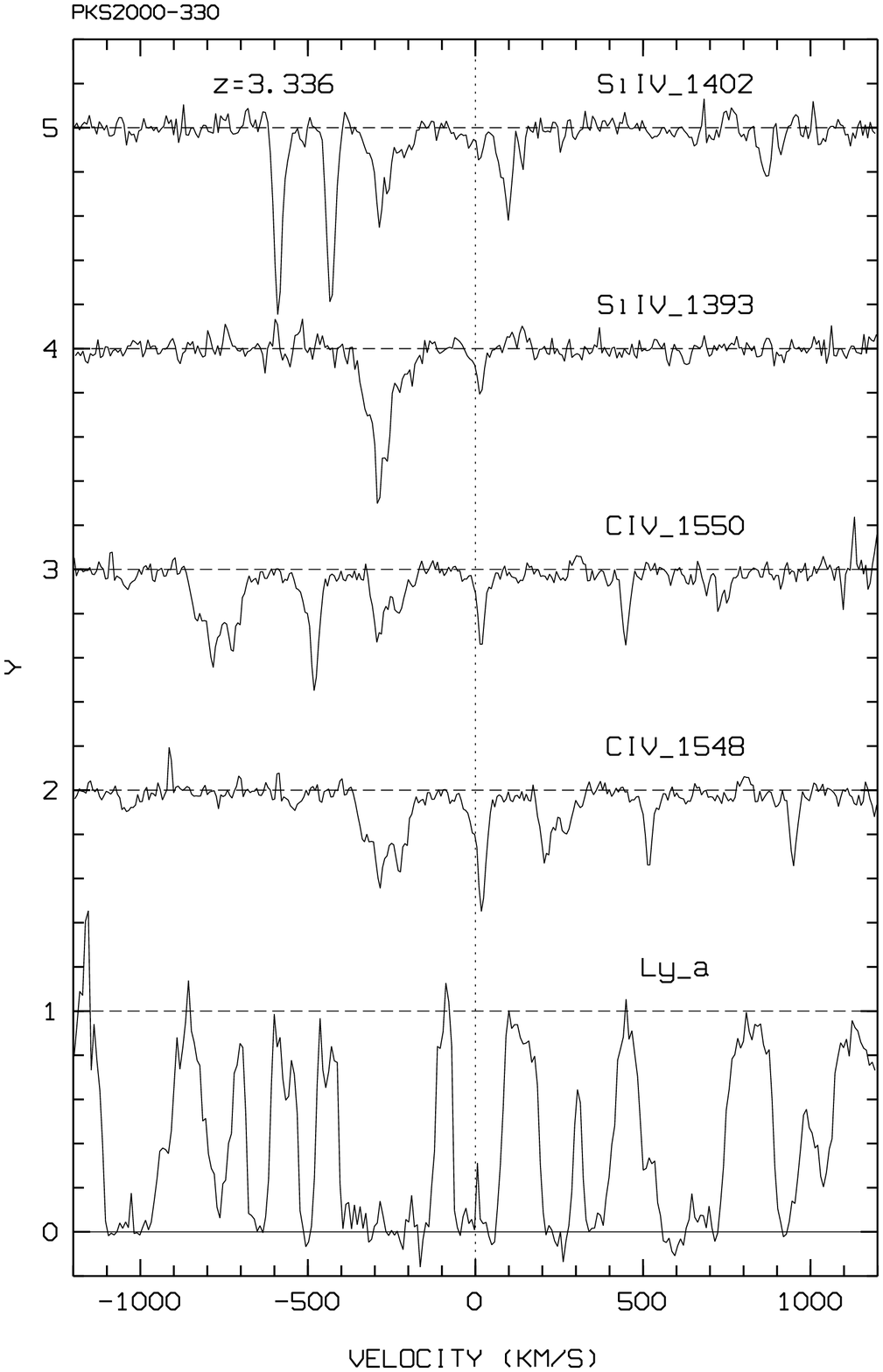,height=16.5cm}
}}
\caption{Cluster of Ly$\alpha$ lines at $z$~$\sim$~3.336 on the line of sight
to PKS~2000-33 on a velocity scale. The corresponding C~{\sc iv} and
Si~{\sc iv} lines are also shown.
}
\end{figure}
\section{The 1D correlation function}
\subsection{The Ly$\alpha$ forest}
If the spatial distribution of the Ly$\alpha$ gas is related to the
mass distribution it is of interest to measure the correlation of the 
absorption lines as a possible probe of the early stages of the gravitational
clustering. Much work has been dedicated to the study of
the 1D clustering properties of the Ly$\alpha$ lines along the line of sight
to the quasars. Till recently no clustering in velocity space had
been detected on scales 300~$<$~$\Delta v$~$<$~30000~km~s$^{-1}$
(Sargent et al. 1980, Bechtold 1987, Webb \& Barcons 1991). Most of the
results were obtained using intermediate resolution spectroscopy. 
The number of lines observed at high resolution and with good S/N ratio
has increased dramatically and it has been possible to investigate the
clustering of the lines for different column density regimes
(Rauch et al. 1992). Cristiani et al. (1995) have found significant 
clustering, with $\xi$~$\sim$~1, at $\Delta v$~=~100~km~s$^{-1}$ for lines 
with log~$N$(H~{\sc i})~$>$~13.8. For log~$N$(H~{\sc i})~$>$~13.3, a 
significant but much weaker signal, $\xi$~$\sim$~0.34, is found. 
This result has been confirmed by Meiksin \& Bouchet (1995) using 
neighbor probability distribution functions. They find also strong
evidence for anticorrelation on the scale of 3--6$h^{-1}$Mpc
(see also Hu et al. 1995). This overall is  
consistent with the idea that strong lines trace the dark matter filaments 
and weak lines are mostly found in underdense regions (Riediger et al. 1996).
\par
\subsection{The metal lines}
It has been shown convincingly that metal line systems at $z$~$<$~1 are 
associated with galaxies (Bergeron \& Boiss\'e 1991, Steidel 1993).
It is thus not surprising to observe that metal line systems do cluster
on scales $\Delta v$~$<$~600~km~s$^{-1}$ (Sargent et al. 1988).
There is however a problem when comparing clustering of absorption lines and
clustering of galaxies.
Indeed when observed at high spectral resolution, metal lines 
break up into individual components. 
The 1D correlation function can be fitted using the sum of two Gaussian
distributions with $\sigma$~=~110 and 525~km~s$^{-1}$ and
$\sigma$~=~80 and 390~km~s$^{-1}$ for the CIV systems at $z$~$\sim$~2.6 
(Petitjean \& Bergeron 1994) and MgII systems at $z$~$\sim$~1 (Petitjean 
\& Bergeron 1990) respectively. The clustering at small velocities 
reflects motions of clouds within one individual halo
whereas larger velocities indicate clustering of halos.\par
It is thus clear that 
there is a difficulty in defining what part of
the correlation function is related to clustering of haloes and
what part is a consequence of motions of clouds within one halo.
Only the first part of the function may be relevant for large scale 
analysis. 
\subsection{The weak C~{\sc iv} systems}
Recent observations have shown that, at $z$~$\sim$~3, C~{\sc iv} is found 
in 90\% of the clouds with $N$(H~{\sc i})~$>$~10$^{15}$~cm$^{-2}$ and in 
about 50\% of the clouds with 
3$\times$10$^{14}$~$<$~$N$(H~{\sc i}) $<$ 10$^{15}$~cm$^{-2}$ (Songaila \&
Cowie 1996, Cowie et al. 1995). Several components are seen in most of these
weak systems thus the correlation function shows a signal on scale
smaller than 200~km~s$^{-1}$. On this basis, Fern\'andez-Soto et al. (1996) 
argue that the observed clustering is broadly compatible with that
expected for galaxies and that most Ly$\alpha$ absorbers arise in galaxies. 
We could argue however, on the basis of the discussion in the previous 
section, that the signal detected in the correlation function has 
nothing to do with clustering of galaxies. What is seen is
just the velocity structure of the Ly$\alpha$ gas inside large complexes. \par
Indeed a more attractive picture arise from the simulations showing that
the Ly$\alpha$ absorption line properties can be understood if the
gas traces the development of structures in the Universe (Cen et al. 1994,
Petitjean et al. 1995, M\"ucket et al. 1996, Hernquist et al. 1996, 
Miralda--Escud\'e et al. 1996). This is illustrated in Fig.~1 showing the
spatial distribution of the Ly$\alpha$ gas in a 2 Mpc slice of
a (25 Mpc)$^3$ simulation box at redshifts $z$ = (3.28, 1.6, 0.8) 
respectively. In this picture, part of the gas is located inside filaments
where star formation can occur very early in small halos that subsequently
merge to build-up a so-called galaxy (Haehnelt et al. 1996). 
This gas contains metals. The remaining part of the gas has no metals and
either is loosely associated with the filaments and 
has $N$(H~{\sc i})~$\ga$~10$^{14}$~cm$^{-2}$
or is located in the underdense regions and has 
$N$(H~{\sc i})~$\la$~10$^{14}$~cm$^{-2}$.
In this picture it might happen that the line of sight intercepts a
filament along its largest dimension. In such a case it is expected to see
a cluster of strong Ly$\alpha$ lines with associated C~{\sc iv} 
lines of very different strengths. Such an observation is shown in Fig.~2. 
The data have been obtained with EMMI at the ESO NTT. The total integration 
time is 18 hours (Petitjean et al. 1996).
\section{The 3D clustering}
Recent studies indicate large scale clustering of absorbers. 
Heisler et al. (1989) detected significant correlation signal 
for C~{\sc iv} systems out to velocities of $\Delta v$ = 10000~km~s$^{-1}$.
Foltz et al. (1993) found an overdensity of C~{\sc iv} systems in the
redshift range 
1.57~$\la$~$z$~$\la$~1.69 along the lines of sight to 6 QSOs.
Francis \& Hewett (1995) discovered two damped Ly$\alpha$ systems at similar
redshifts in two lines of sight separated by 18$h^{-1}$~Mpc comoving.
Most of the time however the signal is a consequence of an unusual 
overdensity of systems along a peculiar line of sight.
Large samples of absorption systems have been used to 
investigate the 3D clustering properties of the absorbers (York et al. 1991,
Tytler et al. 1993) with little success mostly due to lack of data.
\subsection{An example: The field around Q1037-2704}
In this context, the field surrounding the bright ($m_{\rm V}$~=~17.4)
high redshift ($z_{\rm em}$ = 2.193) QSO Tol Q1037-2704
is quite promising. Jakobsen et al. (1986) were the first to note
the remarkable similarity of the metal-line absorption systems in the
spectra of Tol~1037-2704 and Tol~1038-2712 separated by 17'9 on the sky,
corresponding to 4.3$h^{-1}_{100}$~Mpc for $q_{\rm o}$~=~0.5 at $z$~$\sim$~2.
They interpreted
this as evidence for the presence of a supercluster 
along the line of sight to the QSOs. The fact that the number of metal-line
systems in both spectra over the range 1.90~$\leq$~$z$~$\leq$~2.15
is far in excess of what is usually observed has
been considered as the strongest argument supporting this conclusion
(Ulrich \& Perryman 1986, Sargent \& Steidel 1987, Robertson 1987).
In a recent paper, Dinshaw \& Impey (1996) have presented 
new data on four 
quasars in this field. They find that the velocity correlation function
of the C~{\sc iv} systems shows strong and significant clustering 
for velocity separations less than 1000~km~s$^{-1}$ and 
up to 7000~km~s$^{-1}$ respectively. The spatial correlation function 
shows a marginally significant signal on scales of $<$~18~Mpc. 
They conclude that the dimensions of the proposed supercluster are at
least 30~$h^{-1}$~Mpc on the plane of the sky and approximately
80~$h^{-1}$~Mpc along the line of sight. Moreover,
by an analysis of the metal content and ionization state of several C~{\sc iv}
complexes in Q1037--2704, Lespine \& Petitjean (1996) have shown that
the gas lies in intervening systems, supporting the presence of
a coherent structure of supercluster dimensions.\par
\subsection{Clustering of absorbers with VLT}
The best approach to study clustering of absorbers is to search 
a small field for a large number of quasars and to identify the absorption 
systems in the quasar spectra. This will allow 3D mapping of the baryonic 
content of this part of the Universe. 
The major limitation is that the number of 
quasars per square degree is large enough to yield interesting conclusions 
at a magnitude prohibitively large to achieve adequate spectroscopy 
on 4~m class telescopes.\par
The possible VLT project would thus include :\par
-- Deep imaging in broad bands with FORS or EMMI at the NTT to select QSO 
candidates. These images could be used for a parallel
programme to detect galaxies at high $z$.\par
-- Low resolution spectroscopy with FORS to confirm the candidates.
This could be part of a programme aimed at determining the luminosity 
function of AGNs.\par
-- Intermediate resolution ($m_{\rm QSO}$~$<$~22.5) and high 
resolution ($m_{\rm QSO}$~$<$~19) spectroscopy of the QSOs with FUEGOS
and UVES for the brightest to study the absorptions. \par
-- Multi-object spectoscopy in the field with FORS, NIRMOS and ISAAC to
identify the associated galaxies.\par
\begin{table}
\begin{tabular}{c r r r c c c c}
\multicolumn{8}{l}{{\bf Table 1.} One degree field, 
$w_{\rm obs,lim}$~$>$~0.2~\AA}\\
\\
\hline
\multicolumn{1}{c}{\ }&\multicolumn{3}{c}{Nb of QSOs}&
\multicolumn{1}{c}{\ \ \ }&\multicolumn{3}{c}{Nb of absorbers per QSO}\\
\multicolumn{1}{c}{\ }&\multicolumn{1}{c}{$m$~$<$~21}&
\multicolumn{1}{c}{\ \ 22\ \ }&
\multicolumn{1}{c}{\ \ 22.5\ }&\multicolumn{1}{c}{\ \ \ }&
\multicolumn{1}{c}{\ \ \ C~{\sc iv}\ \ \ }&\multicolumn{1}{c}{\ \ \ 
Ly$\alpha$}\ \ \ &
\multicolumn{1}{c}{Mg~{\sc ii}}\\
\hline
\\
~0~$<$~$z$~$<$~2.2& 33 & 74 & 129 && 5 && 0.7 \\
~0~$<$~$z$~$<$~2.2& 7 & 20 & 32   && 7 & 100 & 0.9 \\
\\
\hline
\end{tabular}
\end{table}
Table~1 gives the number of QSOs expected in a one-degree field 
(Hartwick \& Shade 1990) and
the mean number of absorbers, with $w_{\rm obs,lim}$~$>$~0.2~\AA,
per line of sight. These numbers are 
indicative and take into account various observational limitations. 
It is clear that one would like to observe the largest number of QSOs in
the field. A compromize between this number and a reasonable amount of
observing time should be found. However for a random spatial distribution
of the QSOs, a number of hundred QSOs in the field seems adequate
leading to primarily targetting $m$~$<$~22 QSOs. 
The exposure time needed to obtain spectra of S/N ratio of 20 at a 
resolution of $R$~$\sim$~5000 (thus $w_{\rm obs,lim}$~$\sim$~0.2~\AA)
on a $m$~=~22 QSO is about 30~hours. This is large but the use
of the MOS capabilities reduces the effective observing time requested
to achieve the project. In this prospect the instrument providing the 
largest field, thus FUEGOS, should be used. Only six FUEGOS settings will 
be needed.
%
%

%

\begin{thebibliography}
%
\bibitem{}{}{}
Bechtold, J. (1987): {\sl High redshift and Primeval Galaxies}. IAP 
Colloquium, ed. by Bergeron J. et al., Editions Fronti\`eres, Gif sur Yvette,
p. 397 
\bibitem{}{}{} 
Bechtold, J., Crotts, A.P.S., Duncan, C., Fang, Y. (1994): \apjl {\bf 437}, L83
\bibitem{}{}{}
Bergeron, J., Boiss\'e, P. (1991): A\&A {\bf 243}, 344 
\bibitem{}{}{}
Cen, R., Miralda-Escud\'e, J., Ostriker, J.P., Rauch, M. (1994): \apj 
{\bf 437}, L9
\bibitem{}{}{}
Charlton, J.C., Churchill, C.W., Linder, S.M. (1995): \apjl {\bf 452}, L81
\bibitem{}{}{}
Cowie, L.L., Songaila, A., Kim, T.-S., Hu, E.M. (1995): \aj {\bf 109}, 1522
\bibitem{}{}{}
Cristiani, S., D'Odorico, S., Fontana, A., Giallongo, E., Savaglio, S. (1995):
MNRAS {\bf 273}, 1016
\bibitem{}{}{}
Dinshaw, N., Foltz, C.B., Impey, C.D., et al. (1995): Nat {\bf 373}, 223
\bibitem{}{}{}
Dinshaw, N., Impey, C.D. (1996): ApJ {\bf 458}, 73 
\bibitem{}{}{}
Dinshaw, N., Impey, C.D., Foltz, C.B., et al. (1994) \apj {\bf 437}, L87
\bibitem{}{}{}
Fern\'andez-Soto, A., Lanzetta, K.M., Barcons, X. et al. (1996): \apjl {\bf
460}, L85
\bibitem{}{}{}
Foltz, C.B., Hewett, P.C., Chaffee, F.H., Hogan C.J. (1993): \aj 
{\bf 105}, 22
\bibitem{}{}{}
Francis, P.J., Hewett, P.C. (1993): \aj {\bf 105}, 1633
\bibitem{}{}{}
Haehnelt, M.G., Steinmetz, M., Rauch, M. (1996): \apj preprint 
\bibitem{}{}{}
Hartwick, F.D.A., Shade, D. (1990): Ann. Rev. Astron. Astroph. {\bf 28}, 437
\bibitem{}{}{}
Heisler, J., Hogan, C.J., White, S.D.M. (1989): \apj {\bf 347}, 52
\bibitem{}{}{}
Hernquist, L., Katz, N., Weinberg, D.H., Miralda-Escud\'e, J. (1996): 
ApJ {\bf 457}, L51
\bibitem{}{}{}
Hu, E.M., Tae-Sun Kim, Cowie, L.L., Songaila, A., Rauch, M. (1995): \aj
{\bf 110}, 1526 
\bibitem{}{}{}
Jakobsen, P., Perryman, M.A.C. (1992): \apj {\bf 392}, 432
\bibitem{}{}{}
Jakobsen, P., Perryman, M.A.C., Ulrich, M.H., et al. (1986): \apj 
{\bf 303}, L27
\bibitem{}{}{}
Lanzetta, K.M., Bowen, D.V., Tytler, D., Webb, J.K. (1994): ApJ {\bf 442}, 538
\bibitem{}{}{}
Le Brun, V., Bergeron, J., Boiss\'e, P., 1996, A\&A {\bf 306}, 691
\bibitem{}{}{}
Lespine, Y., Petitjean, P. (1996): A\&A in press
\bibitem{}{}{}
Meiksin, A., Bouchet, F. (1995): \apjl {\bf 448}, L88
\bibitem{}{}{}
Miralda-Escud\'e, J., Cen, R., Ostriker, J.P., Rauch, M. (1996): \apj in press
\bibitem{}{}{}
M\"ucket, J., Petitjean, P., Kates, R.E., Riediger, R. (1996): A\&A 
{\bf 308}, 17
\bibitem{}{}{}
Petitjean P., Bergeron J. (1990): \aeta {\bf 231}, 309
\bibitem{}{}{}
Petitjean P., Bergeron J. (1994): \aeta {\bf 283}, 759
\bibitem{}{}{}
Petitjean, P. (1995): {\it Science with VLT}.
ESO Workshop, ed. by Danziger J., Walsh J., Springer, Heidelberg, p. 339
\bibitem{}{}{}
Petitjean, P., M\"ucket, J., Kates, R.E. (1995): \aetal {\bf 295}, L9
\bibitem{}{}{}
Petitjean, P., Fontana, A., Giallongo, E., Lespine, Y. (1996): in preparation
\bibitem{}{}{}
Rauch, M., Haenelt, M. (1996): \mn {\bf 275}, L76
\bibitem{}{}{}
Rauch, M., Carswell, R.F., Chaffee F.H. et al. (1992): \apj {\bf 390}, 387
\bibitem{}{}{}
Riediger, R., Petitjean, P., M\"ucket, J. (1996): in preparation
\bibitem{}{}{}
Robertson, J.G. (1987): \mn {\bf 227}, 65
\bibitem{}{}{}
Robertson, J.G., Shaver, P.A. (1983): \mn {\bf 204}, 69P
\bibitem{}{}{}
Sargent, W.L.W., Young, P.J., Boksenberg, A., Tytler, D. (1980):
\apjs {\bf 42}, 41
\bibitem{}{}{}
Sargent, W.L.W., Steidel, C.C. (1987): \apj {\bf 322}, 142
\bibitem{}{}{}
Sargent, W.L.W., Boksenberg, A., Steidel C.C. (1988): \apjs {\bf 68}, 539
\bibitem{}{}{}
Shaver, P.A., Robertson, J.G. (1983): \apjl {\bf 268}, L57
\bibitem{}{}{}
Smette, A., Robertson, J.G., Shaver, P.A., Reimers, D., Wisotzki, L., 
K\"ohler, T. (1995): \aetas {\bf 113}, 199
\bibitem{}{}{}
Smette, A., Surdej, J., Shaver, P.A., et al. (1992): \apj {\bf 389}, 39
\bibitem{}{}{}
Songaila, A., Cowie, L.L. (1996): \aj preprint
\bibitem{}{}{}
Steidel C.C. (1993): {\sl Third Tetons Summer School,
The Environment and Evolution of Galaxies}, ed. by J.M. Shull and H.A. 
Thronson Jr., Kluwer, Dordrecht, p. 263
\bibitem{}{}{}
Tytler, D. et al. (1993): ApJ {\bf 405}, 57
\bibitem{}{}{}
Ulrich, M.H., Perryman, M.A.C. (1986): \mn {\bf 220}, 429
\bibitem{}{}{}
Webb, J.K., Barcons, X. (1991): MNRAS {\bf 250}, 270
\bibitem{}{}{}
Wolfe, A.M. (1993): \apj {\bf 402}, 411
\bibitem{}{}{}
York et al. (1991): MNRAS {\bf 250}, 24
%
\end{thebibliography}
\end{document}